# Authorship-contribution normalized Sh-index and citations are better research output indicators

Vishvesh Karthik[1], Indupalli Sishir Anand[1], Utkarsha Mahanta[1], and Gaurav Sharma[1*]

[1] Department of Biotechnology, Indian Institute of Technology Hyderabad, Sangareddy, Telangana, India

\* Corresponding authors: sharmaG@bt.iith.ac.in

**Abstract**: Bibliometric measures, such as total citations and h-index, have become a cornerstone for evaluating academic performance; however, these traditional metrics, being non-weighted, inadequately capture the nuances of individual contributions. To address this constraint, we developed GScholarLens, an open-access browser extension that integrates seamlessly with Google Scholar to enable detailed bibliometric analysis. GScholarLens categorizes publications by authorship roles, adjusts citation weightings accordingly, and introduces Scholar h-index (Sh-index), an authorship-contribution normalized h-index. This tool proportionally weights citations based on authorship position using heuristic percentages, i.e., corresponding 100%, first 90%, second 50%, co-authors in publications with ≤6 authors 25%, and co-authors in publications with ≥6 authors 10%. As of now, there is no empirical data available for author-contribution weights, however, this proof-of-concept framework can easily adapt more precise author-contribution weightage data decided by authors at the time of manuscript submission along with CRediT (Contributor Roles Taxonomy), which can be mandated by journals and publishers. Furthermore, this tool incorporates retraction detection by mapping data from retraction databases into the Google Scholar interface. By aligning bibliometric evaluation more closely with actual scholarly contribution, GScholarLens presents a better open-access framework for academic recognition, particularly within interdisciplinary and highly collaborative research environments.

This tool is freely accessible at https://project.iith.ac.in/sharmaglab/gscholarlens/.

**Keywords:** Publishing, citations, h-index, Sh-index, academia, researchers, collaborations

**Main**

With the rise of internet, open-accessibility, and data-driven tools, bibliometric metrics such as publication citations, h-index, and journal impact factor are being generously used to evaluate the performance of any scholar by researchers in all fields (Hicks et al., 2015; Pasterkamp et al., 2016). The unfortunate reality of academia is that irrespective of higher-ranked officials saying they don't care about these metrics, they are always looking for the best publication, where it was published, how many citations it has as of today, were you the corresponding/lead/co-author, and what is the h-index of the researcher, during faculty selection, evaluation, and promotion. Even an early-stage researcher or student judges the quality of a PI or higher-ranked academic official based on the above mentioned in their CV, personal/professional webpages, openly-available and user-maintained academic search engines and bibliographic databases (ASEBDs) such as Google Scholar (Gusenbauer, 2019), or subscription based tools such as, Scopus, Web of Science, etc.

Being cumulative in nature, these traditional metrics fail to differentiate the varied contributions of individual author in each publication. Other metrics such as Author Impact Factor (AIF) have been introduced to capture such temporal variations but they diverge from simplicity (Pan and Fortunato, 2014). Take an example of two researchers, who have been working and publishing together exclusively; their citations and h-index will be the same. However, one of them can be the main workforce of the project, i.e., PI or the first author, and another can be just a coauthor. Even amidst a PI and first author, the contributions can vary significantly; for example, the contributions of a PI can range from brainstorming for months/years to design the experimental/computational/theoretical project, trying to secure funding for years, guiding all authors and collaborators, interpreting results, writing/editing the first draft to final submission. For a first author (PhD or postdoc), the responsibilities can be working on the project for their major thesis/project work, reviewing the literature, analyzing methodology, interpreting results thoroughly, developing newer skills, and writing the early drafts. For a coauthor, the contributions usually start in the middle or partial end of the project, where they will be contacted for their services to wrap up a story in an interdisciplinary research manner. Overall, research teams are often diverse in terms of expertise and size, which complicates the fair distribution of credit. Therefore, the modern collaborative research requires better tools that can evaluate scholarly contributions in an open access user-friendly manner (Wilsdon, 2015).

To address these challenges in a user-friendly manner, we have created **GScholarLens**, an open-access browser-extension designed to operate simple calculations that can be mined from and integrated into the most-accessible ASEBDs, i.e., Google Scholar. This extension tool can be easily installed using this link: https://project.iith.ac.in/sharmaglab/gscholarlens/, which will get automatically incorporated into the Google Scholar page in the same browser. GScholarLens is built upon the fact that citations and metrics are not merely indicators of scholarly impact but also reflects underlying intellectual contributions. Following this ideology, GScholarLens scans the Google Scholar profile, categories primary and subsidiary contributions by researchers and enables proportional distribution of citation and author-level metrics based on their respective contributions in each publication.

This browser extension integrates seamlessly with the Google Scholar interface, functioning on both Google Chrome and Mozilla Firefox platforms. It classifies publications according to authorship categories, i.e., first, second, coauthor, and corresponding, as listed in the Google Scholar profile, which is editable by the individual. It simultaneously stratifies these categories based on Scopus quartile journal rankings (Q1, Q2, Q3, Q4, and NA), suggesting the putative journal quality scores. The tool further organizes citations data into authorship and quartile categories, enabling granular analysis of citation patterns. A key feature of this extension is its visualization of citation distributions for each authorship category using violin plots, which provide statistical summaries including median, mean, minimum, maximum, and interquartile ranges (25% and 75%). Later, the tool computes and displays the relative percentage contributions of each authorship category to both publication counts and citations, offering a proportional representation of authorship and citation dynamics in a relative manner. Another key functionality of this extension is year-wise categorization of these statistics, allowing users to analyse contributions in any specific year, year range, and cumulatively till any year of research career, where the starting year is automatically calculated based on the first publication reported in the respective Google scholar profile.

This tool introduces a normalized variant of the h-index, termed the **<u>S</u>cholar <u>h</u>-index or (Sh-index)**. To calculate Sh-index, this tool first adjusts citation counts based on authorship roles, assigning proportional weightages as follows: 100% for corresponding authors, 90% for first authors, 50% for second authors, and 10–25% for co-authors depending on the total number of contributors to the publication. We recognize a group of six researchers as an apt-size team to do interdisciplinary collaborative research in an ideal collaborative manner. Co-authors in publications with ≤6 authors are assigned a 25% weight, whereas those having publications with ≥7 authors receive a 10% weight per co-author. Later, these normalized citation values are computed for each publication linked to a researcher's Google Scholar profile, and the Sh-index is subsequently derived using the standard h-index methods. Additionally, the tool provides h-indices disaggregated by authorship categories, offering a comprehensive perspective on scholarly contributions across different roles.

Despite the growing concerns for retracted publications, Google Scholar does not provide information about all retracted papers for individual researcher. To do so in an effective manner, GScholarLens maps all publication titles to Retraction database (Oransky, 2018), providing information about retracted papers directly on the Google Scholar profile as total numbers on the upper panel and later as red-font flag for each article. This strategy allows the viewer to get a comprehensive understanding about the researcher's retracted papers' profile on the Google scholar page itself.

GScholarLens also provides comprehensive statistics such as the median number of raw citations, median adjusted citations, count of publications having zero citations, total retractions, and total preprints available in the profile. Moreover, the tool displays a bar plot describing how many papers the author has published over the past ten years; therefore, offering a clear visualization of the author's annual publication rate. Such visualization facilitates the assessment of high-frequency publication behaviour, potentially spotting authors who are publishing like a paper mill.

It must also be mentioned that we are utilizing Google Scholar owing to its wider audience and editable nature and we are not affiliated to it in any form. The erratic representation of authorship across journals poses a significant challenge for its accurate detection. In this context, Google Scholar's editable nature by the respective researcher offers a practical advantage for tools like GScholarLens. Therefore, to use GScholarLens efficiently and accurately, it is advisable to consider the Google Scholar profile as an online curriculum vitae having publications, which must be updated regularly with proper authorship roles. As any data analytics tool, GScholarLens works on the assumption that identifiers such as * (for all corresponding authors) and ^ (for all first authors) are used in author name formats to deconvolute author contributions properly; if no symbols are used, only the first and last author will be considered as first and corresponding author, respectively. Please have a look at other disclaimers on the GScholarLens website.

Journals have been using CRediT author statement for several years now, which aims to acknowledge individual author contributions, minimize authorship disputes, and promote effective collaboration. However, CRediT information is not being correlated with the matrices and indexes for academic evaluation. We acknowledge that the percentage weights assigned to author positions in our contribution-normalized Sh-index are heuristic and not derived from empirical evidence of actual contributions. These values were chosen to reflect broadly accepted conventions in many scientific fields, where first and corresponding authors are often considered primary contributors and other positions have performed lesser amount of work in the respective publication. We also recognize that individual cases may deviate substantially such as in some or many cases, middle authors might be contributing more than what positional proxies are being used in this tool. Thus, the present model should be regarded as a proof-of-concept framework rather than a definitive standard. The long-term feasibility of such contribution-aware metrics will depend on the systematic collection of structured contribution data, such as journal-mandated CRediT statements or percentage-based author declarations. We suggest that journals should ask percentage-based contribution for each author as a declaration along with journal-mandated CRediT statements. For data analysis and metrics calculation purpose, such weightage contribution should be kept in the metadata of Digital Object Identifier (DOI), like affiliations. Our proposed framework is therefore intended to stimulate discussion and highlight the need for incorporating contribution transparency into bibliometric evaluations, rather than to prescribe fixed weight assignments.

By offering a transparent, data-enriched platform for interrogating the relationships between authorship-based contributions and weighting their citations per publication, GScholarLens aims to advance discussions about integrity in academic recognition systems. Through a robust integration of bibliometric analysis and qualitative assessments, GScholarLens fills up the lacunae in the conventional measures of scholarly impact. Owing to its contribution weightage-based normalization, Sh-index can provide better insights about the research contribution of any researcher. Eventually, GScholarLens, via its authorship-based productivity and Sh-index calculations, can empower researchers, institutions, and policymakers to adopt more comprehensive and just systems for evaluating scholarly work, fostering academic culture where contributions of all forms are acknowledged and celebrated.

**Ethics approval and consent to participate:** Not applicable

**Availability of data and materials:** Authors have used open-source tools in this analysis. The information of all used tools and their versions have been provided on the GScholarLens website https://project.iith.ac.in/sharmaglab/gscholarlens/. GScholarLens extension can be installed using these links, following which, this tool sign will be automatically visible on the Google Scholar page itself:

Chrome: https://chromewebstore.google.com/detail/scholarlens/ehmbblmfmfbnilannaiadknogghebnph

Mozilla: https://addons.mozilla.org/en-US/firefox/addon/scholarlens/

**Competing interests:** The authors declare no conflict of interest to disclose.

**Funding:** GS acknowledges the seed grant from IIT Hyderabad and Start-up Research Grant (SRG) from Science and Engineering Research Board (SERB) for supporting his research.

**Authors' contributions:** GS conceptualized the idea and supervised the project. VK and ISA performed the analysis. UM did the preliminary work for this tool's concept note. GS and VK wrote, edited and finalized the manuscript. All authors approved the final version.

**Figure Legend:**

**Figure 1: Screenshots illustrating the visual representation of various analytical outputs created by GScholarLens. a)** Visualization of Sh-index and h-index metrics across different authorship categories, accompanied by diverse statistics including the number of retracted papers, preprints, and publications with zero citations. The right panel displays publication counts per year, facilitating the identification of researchers with unusually high publication frequencies, potentially indicative of 'paper mill' or highly collaborative activity. **b)** Aggregated publication and citation count categorized by authorship category and Scopus journal quartile rankings (Q1-Q4, and NA), represented via color gradient for differentiation. **c)** The upper panel shows the violin plot for citations based on authorship category on a log scale. The bottom panels depict publication count and citation-wise contribution based on authorship categories out of 100. **d)** Each of the retracted articles, as shown in panel (a), is distinctly marked with a "**RETRACTED**" badge, allowing for immediate visual identification of such records. **e)** Basic bibliometric information, i.e., total citations and h-index, as retrieved directly from Google Scholar.

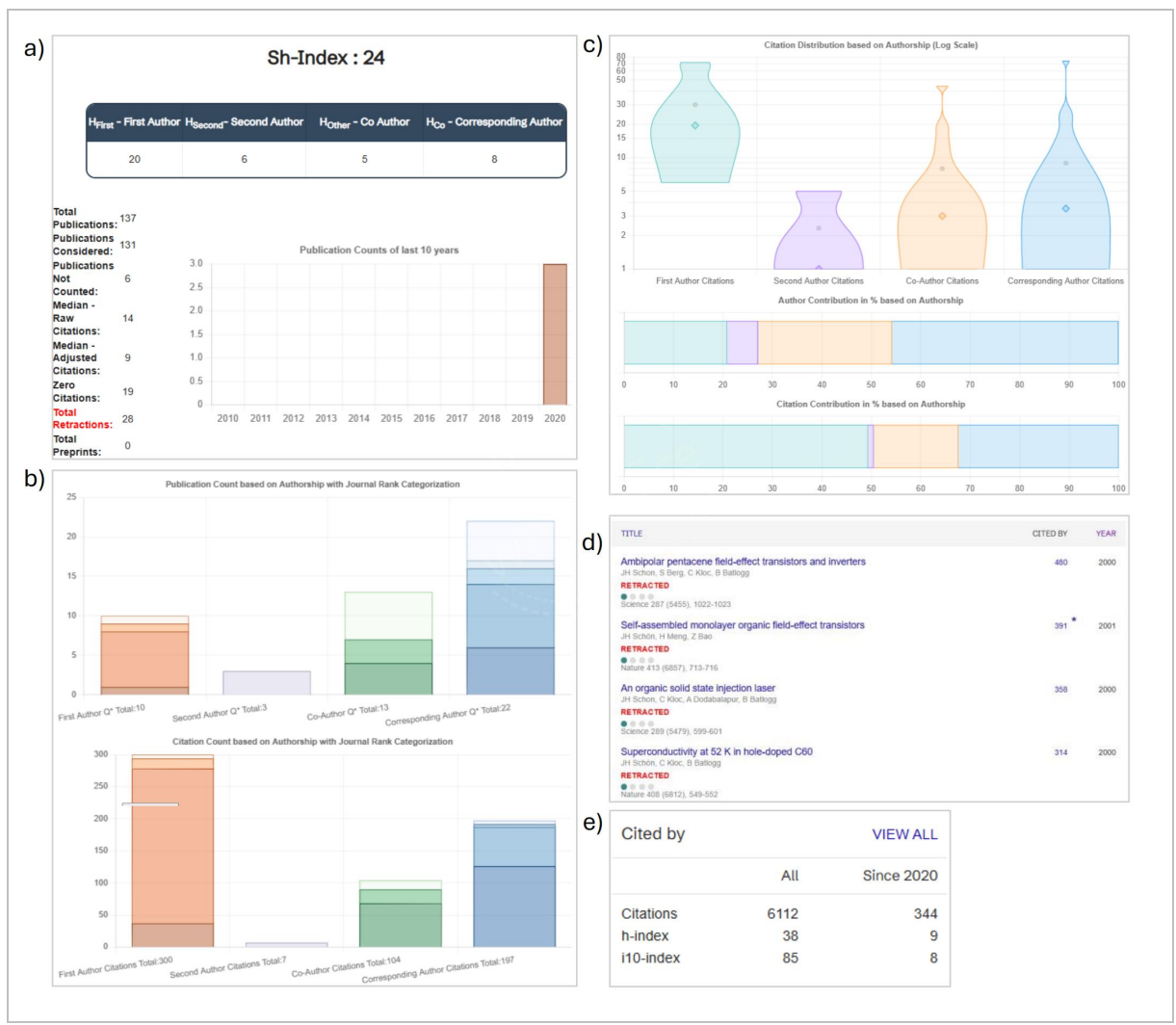